\documentclass[a4paper]{jpconf}
\usepackage{graphicx}
\begin{document}

\title{Apart Cen A are UHECR   mostly \\ heavy radioactive and  galactic  nuclei?
  }
\author{Daniele Fargion}

\address{ Physics Department, Rome University 1,INFN Rome1, Ple. A. Moro 2, 00185}

\ead{daniele.fargion@roma1.infn.it}

\begin{abstract}
Earliest AUGER (the Pierre Auger Observatory) UHECR (Ultra High Energy Cosmic Rays) anisotropy correlated with AGN (active galactic nuclei) within a  GZK (Greisen-Zatsepin-Kuz'min  cut off)  Universe almost fade away. Recent UHECR mass compositions did show a  negligible nucleon composition and an UHECR nuclei (light or heavy) signature. Indeed  last map miss the Super-Galactic Plane. The absence of  UHECR  events toward the Virgo cluster, an unique spread clustering of events around  Cen-A, our nearest AGN, suggested a  He-like  nuclei as the main extragalactic UHECR component from Cen A, coexisting with Auger and HIRES  (High-Resolution Fly's Eye) composition. Because the light nuclei fragility  such He UHECR cannot arrive from Virgo (being too far). Multiplet at twenty EeV along Cen A recently discovered by Auger confirm this interpretation as being foreseen to be indebt to fragments ($D,He^{3}, p$) that had to reach us along the same UHECR. However  remaining majority of UHECR clustering are partially correlated  with a gamma noise at (1-3 MeV) in Comptel sky, linked to $Al^{26}$ galactic radioactive map as well as to a few TeV gamma (ICECUBE-ARGO) anisotropy maps; rare UHECR triplet are overlapping on Vela  TeV anisotropy and other nearest galactic gamma sources (as partially Crab and a Galactic core corona). Therefore UHECR   might be also (or mostly) heavy radioactive galactic nuclei  as $Ni^{55}$, $Ni^{56}$,  $Ni^{57}$  and $Co^{57}$ bent from the sources whose $\beta$ and $\gamma$ radioactivity and decay in flight is boosted (by Lorentz factor $\Gamma_{Ni}\simeq 10^{9}- 10^{8}$),  leading to  TeV correlated sky anisotropy. Galactic UHECR signals inside the inner center maybe suppressed by the largest spreading repulsive Lorentz bending forces. More clustering around external galactic plane is nevertheless expected in present and future data. Magellanic Cloud and Magellanic  Stream may also rise more and more in UHECR maps (as well as in multiplet signals). Future UHECR clustering might be observed  around Cas A and Cygnus by T.A. (Telescope Array). The UHECR spectra cut off may be not an extragalactic GZK feature but just the imprint of a galactic confinement and-or  a spectroscopic heavy composition decrease step.
\end{abstract}

\section{Cen A and multiplet fragments versus heavy radioactive nuclei decay in flight}
\begin{figure}[htb]
\begin{center}

\includegraphics[width=3.1 in]{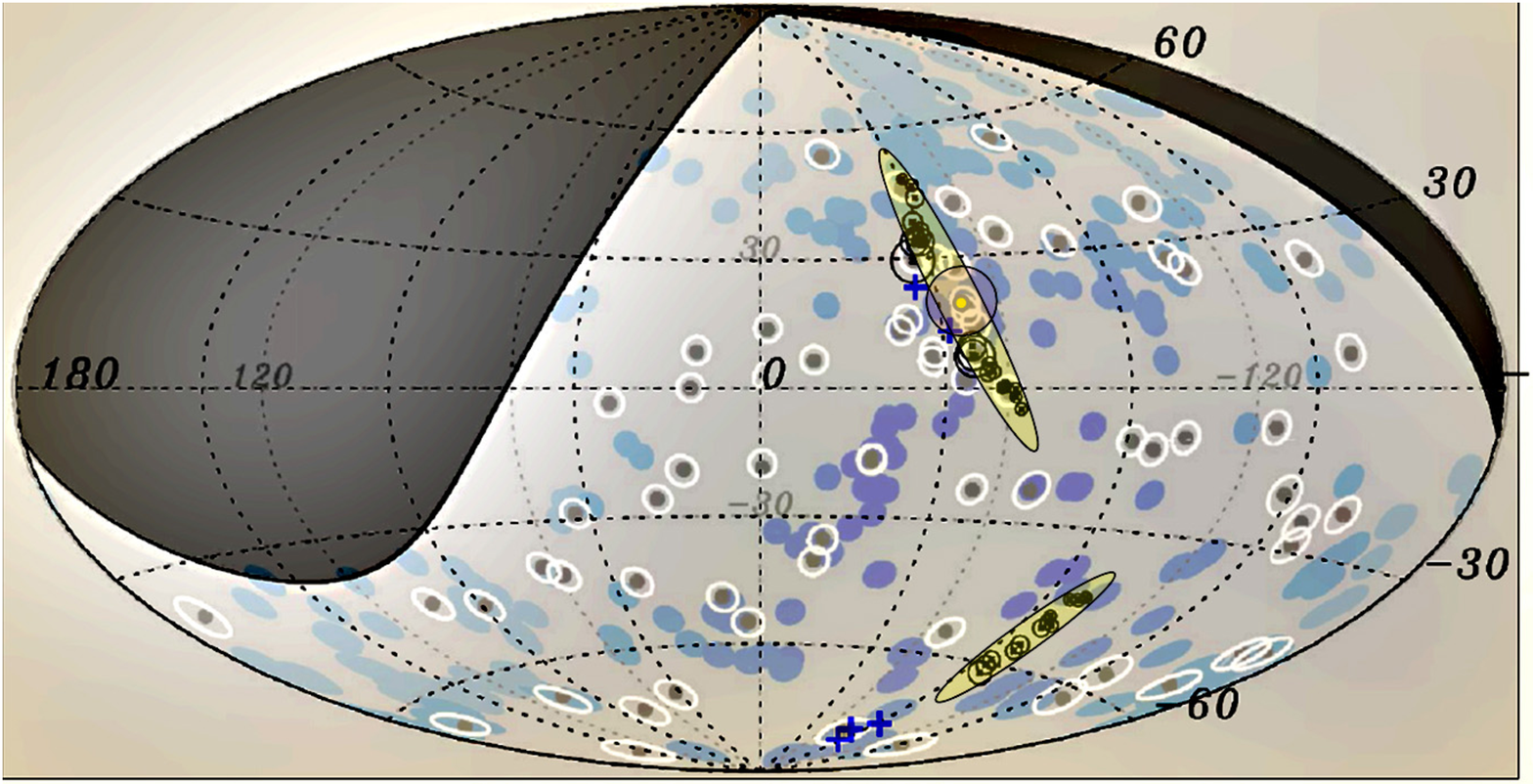}
 \includegraphics[width=3.1 in]{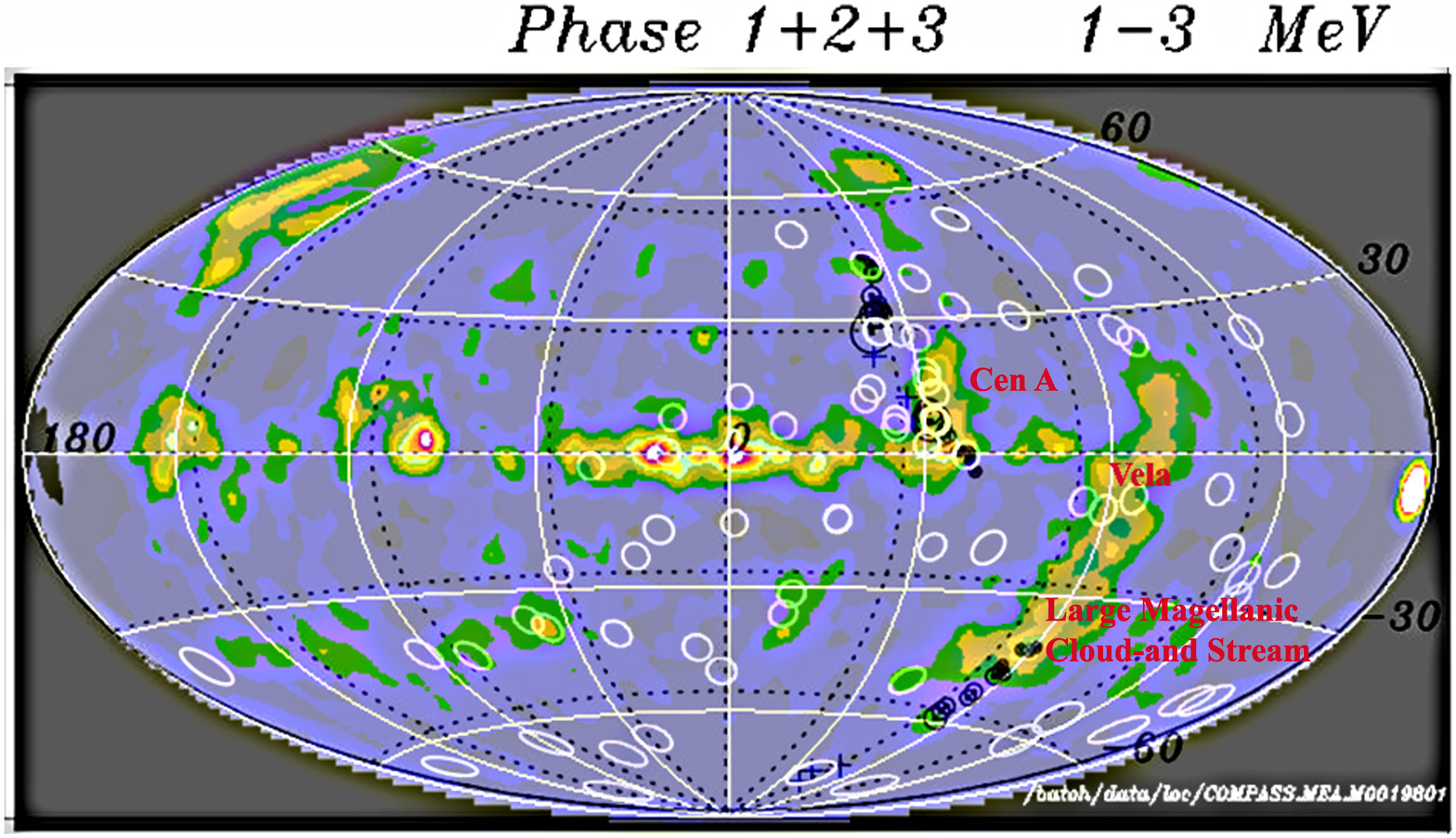}
\caption{ Left: The AUGER UHECR event map and two of the three AUGER multiplet clustering toward Cen A; a thin narrow elliptical area and a small disk mark the place.  A third Multiplet clustering points toward Large and Small Magellanic clouds and it overlaps on Magellanic stream. Right: last UHECR event map by AUGER where the clustering near Cen A overlap the MeV Comptel gamma (1-3 MeV)  map; note connected map beside nearby AGN Cen A, along the nearest pulsar Vela, the Magellanic stream and the galactic core. }\label{1r}
\end{center}
\end{figure}

\begin{figure}[!t]
  \vspace{5mm}
  \centering
  \includegraphics[width=4.8 in]{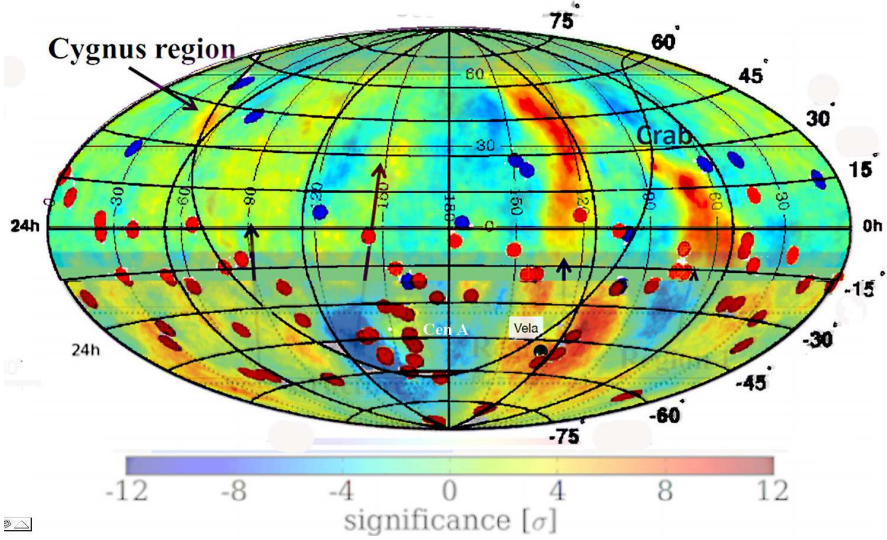}
  \caption{
  The AUGER 2010 UHECR (red) and Hires (blue) event map in celestial coordinate on recent (2010) TeV diffused CR map (ARGO-Milagro-ICECUBE) and labels. See \cite{ARGO},\cite{Desiati}, and references. The triplet UHECR event clustering toward Vela and the TeV spread activity around  is remarkable. The Cen A clustering at the fig. center is the main feature in the UHECR map. ARGO TeV anisotropy born around Crab connect and overlap the UHECR events in nearby Orion TeV region. Note doublet along the galactic plane and the  TeV near Cygnus and Cas A regions, where Hires did and we foresee TA may detect UHECR events.}
  \label{fig4-5}
 \end{figure}

Ultra High Energy Cosmic Rays, UHECR, Astronomy and nuclear composition are more and more in severe conflict. This contradiction was already inscribed in  early apparent discover of a Super Galactic UHECR correlation \cite{Auger-Nov07} with a first UHECR nuclei composition signature. It was note realistic for UHECR to arrive exactly  from Super Galactic plane, as apparent map were suggesting, while being heavy nuclei (and not nucleon).
 A very heavy UHECR composition being greatly bent by galactic magnetic fields cannot open to any sharp astronomy nor  correlate to Super Galactic Plane, nor it can explain easily the unique observed mild UHECR Cen A event clustering.     Heavy nuclei as Fe, Ni, Co   may  however induce large scale anisotropy or spread inhomogeneity (as possibly from a nearby source, Vela). Moreover let us remind that  UHECR (if nucleon and if  extragalactic) are making photo-pion interacting on cosmic CMBR (Cosmic Microwave Background Radiation) and a consequent EeV gamma and neutrinos secondaries. These expected signals at EeV for gamma and tau neutrinos are more and more constrained being  at the edge of AUGER detection, see \cite{Auger08}.  If UHECR are light nuclei they may produce  fragment secondaries nuclei or nucleon, as well as gamma-ray and neutrino tails at tens of TeV-PeV because nuclei photo-dissociation; also heavy nuclei, if extragalactic, may lead by photodisintegration on extragalactic flight leading to such TeVs-PeV gamma and neutrino tail. If UHECR are very heavy  they maybe bent and bound and confined within their local birth source group. For instance UHECR heavy nuclei maybe bent and confined within Virgo group ; in analogy heaviest UHECR may be bent, crowded and nearly contained within our galaxy, but they cannot produce much photodisintegration secondaries. However \emph{UHECR radioactive nuclei} as  $Ni^{56}$, $Ni^{57}$ and $Co^{57}$ $Co^{60}$, while flying and decaying, may trace their presence by boosted gamma (and positron) tails and beta decay neutrinos. If such UHECR are produced by Super-novae remanent, SNRs (SuperNova Remanents) , or better by their twin micro-quasars jets (in GRB-SN asymmetric jet explosions and late precessing jet model, see \cite{Fargion1998}) these tails might be observable. The relic gamma radiation (in rest frame of the nuclei around hundred keV) will be boosted by UHECR $E \geq 6 \cdot 10^{19} eV$ energy, heavy nuclei huge Lorentz factor ($\Gamma_{Ni^{55}} \simeq 10^{9}$), shining at our laboratory system around tens-hundred TeV gamma region, observable at Milagro-ARGO-ICECUBE  TeVs CR maps as their correlated anisotropy, see Fig \ref{fig4-5}. A comparable trace somehow  correlated, but at more non relativistic stages, is offered by heavy nuclei (as  the $Al^{26}$ at MeV Comptel gamma maps see Fig. \ref{1r}) in galactic plane  \cite{Fargion2011}. Inspired by this  first radioactive gamma-UHECR connection possibility   now we extend such understanding  for the partial gamma TeV-UHECR connection anisotropy assumed born by  boosted nuclei decay, see Fig.\ref{fig4-5}. As a natural consequence present UHECR cut off may be related not to any real nucleon extragalactic GZK cut-off but  to a  heavy nuclei spectroscopy step, a confinement in our own galaxy, an  nearby anisotropic distances dilution. The same heavy nuclei ejection by nearby Supernova maybe the source of local nuclei anisotropy made by asymmetric SN-jet explosions (see:\cite{Hansen:SN-Uranium}).

   Extragalactic UHECR around CenA  formed (mostly) by  lightest nuclei may explain a partial clustering of events, and the UHECR absence  around Virgo. Light nuclei are fragile and fly few Mpc before being  halted by photo-disruption \cite{Fargion2008},\cite{Fargion2009},\cite{Fargion2010}. The fragments $He + \gamma \rightarrow D+D,D+\gamma \rightarrow p+n+ \gamma  , He + \gamma \rightarrow He^{3}+n, He + \gamma \rightarrow T +p $  may nevertheless trace the same UHECR maps by a secondary  clustering at half or even fourth of the UHECR primary  energy \cite{Fargion09a}, \cite{Fargion2011}. At lower energy (at ten EeV or below) the huge smeared cosmic ray isotropy and homogeneity may hide these tiny inhomogeneity traces.  Nevertheless the very recent AUGER EeV map did not show the expected solar Compton Getting anisotropy but an anisotropy in a different region.
   This result partially disagree with a (widely accepted) extragalactic EeV UHECR component but it favors a more local (galactic) EeV  UHECR presence whose smearing is related to the large Lorentz angle bending and whose non-Compton-Getting inhomogeneity is related to nearby sources, as we are also suggesting in present article.

\section{TeVs gamma near Vela: Boosted and bent $Ni^{55}$-$Co^{60}$ decaying nuclei?}

      Since earliest maps we found that Cen A (the most active and nearby extragalactic AGN) is apparently shining  UHECR source whose clustering (almost a quarter of the event) along a narrow solid angle around (whose opening angular size is  $\simeq 17^{o}$) seem  firm and it is  favoring as we mentioned, lightest nuclei \cite{Fargion2008},\cite{Fargion09a}, \cite{Fargion09b}, \cite{Fargion2009}. But what is the nature and origin of all the other UHECR events? Composition favor mostly  heavy nuclei and as we shall see in figure below by local sources.
       In recent maps of UHECR we noted first hint of galactic source   rising as an UHECR triplet \cite{Fargion09b}.  Also the lesson of $Al^{26}$ and 1-3 MeV gamma map traced by Comptel somehow overlapping with UHECR events favors a role of UHECR radioactive elements.   The most prompt ones are the  $Ni^{55}$, $Ni^{56}$ (and $Co^{56}$ ) made by Supernova (and possibly by their collimated GRB micro-jet components see \cite{Fargion1998}) ejecta in our own galaxy.  Indeed in all SN Ia models as well in SN II (as Vela and Crab probable explosions), the decay chain  $Ni^{56}\rightarrow Co^{56} \rightarrow Fe^{56}$
provides the primary earliest source of energy that powers the supernova optical display. The $Ni^{56}$ decays by electron capture and the daughter $Co^{56}$  emits gamma rays by the nuclear de-excitation
process; the two characteristic gamma lines are  respectively at $E_{\gamma} =$ 158 keV and $E_{\gamma}=$ 812 keV. Their Half lifetime are spread from $35.6$ h for  $Ni^{57}$ and $6.07$ d. for   $Ni^{56}$. However there are also more unstable radioactive rates as  for $Ni^{55}$ nuclei whose half life is just $0.212$ s or $Ni^{67}$ whose decay is $21$ s. Therefore we may have an apparent boosted  ($\Gamma_{Ni^{56}} \simeq 10^{9}$) life time spread from $2.12 \cdot 10^{8}$ s or $6.7$ years (for $Ni^{55}$) up to nearly  $670$ years (for $Ni^{67}$) or $4$ million years for  $Ni^{57}$. This consequent wide range of lifetimes guarantees a long life activity on the UHECR radioactive traces. However the brightest are often the fast decaying ones. The arrival tracks of these UHECR radioactive heavy nuclei may be  widely bent, as shown below, by galactic magnetic fields. Among the excited nuclei to mention  for the UHECR-TeV connection is $Co_{m}^{60}$  whose half life is $10.1$ min and whose decay gamma line is at $59$ keV. At a boosted nominal Lorentz factor $\Gamma_{Co^{60}}= 10^{9}$ we obtain $E_{\gamma}\simeq 59 $ TeV ; its boosted lifetime is $19000$ years or 6kpc distance. Therefore $Co_{m}^{60}$ energy decay traces and lifetime fit well within  present UHECR-TeV connection for nearby galactic sources as Vela. Other radioactive beta decay, usually at higher energy may also shine at hundred or tens TeV or below by inverse Compton and synchrotron radiation. Let us now consider the UHECR and Lorentz bending for  $He_{2}$ around Cen A and $Ni_{28}$, $Co_{27}$ charges.
Cosmic Rays are blurred by magnetic fields. Also UHECR suffer of a Lorentz force deviation. This smearing maybe source of UHECR features. Mostly along Cen A where a narrow UHECR tails, possibly being light nuclei are bent by Lorentz forces. There are  at least three main spectroscopy of UHECR along galactic plane made by galactic fields.  A late nearby (almost local) bending by a nearest coherent galactic arm field, a random one by turbulence and a random one along the whole plane inside different arms: (1)The coherent Lorentz angle bending $\delta_{Coh} $ of a proton (or nuclei) UHECR  within a galactic magnetic field, assuming  a final flight nearby coherent length  of $l_c = 1\cdot kpc$ is $ \delta_{Coh-p} .\simeq{2.3^\circ}\cdot \frac{Z}{Z_{H}} \cdot (\frac{6\cdot10^{19}eV}{E_{CR}})(\frac{B}{3\cdot \mu G}){\frac{l_c}{kpc}}$.
  This value is  quite different for a heavy nuclei as an UHECR from Vela whose distance is only $0.29$ kpc:
 $$\delta_{Coh-Ni} \simeq
{18,7^\circ}\cdot \frac{Z}{Z_{Ni^{28}}} \cdot (\frac{6\cdot10^{19}eV}{E_{CR}})(\frac{B}{3\cdot \mu G})({\frac{l_c}{0.29 kpc})}$$
  Note that this spread is  able to explain the nearby Vela TeV anisotropy area around its correlated UHECR triplet.
 $$\delta_{Coh-Ni} \simeq
{129^\circ}\cdot \frac{Z}{Z_{Ni^{28}}} \cdot (\frac{6\cdot10^{19}eV}{E_{CR}})(\frac{B}{3\cdot \mu G})({\frac{l_c}{2 kpc})}$$
  Note that this spread might be also able to explain the localized TeV anisotropy born a bit far from Crab (2 kpc) apparently  extending  around  a far TeV area near Orion, overlapping also near Gum Nebulae, where spread UHECR events seem to be clustered.(2) The random bending by random turbulent magnetic fields, whose coherent sizes (tens parsecs) are short and its final deflection angle is much smaller than others, here it is ignored.
(3) The ordered aligned and vertical (respect the galactic plane) multiple UHECR bending while skimming along the galactic plane and crossing alternate spiral arm (magnetic field directions): its final random deflection angle is remarkable and discussed below. This random  bending along the galactic plane and arms, $\delta_{rm} $, is created while crossing  the whole Galactic disk $ L\simeq{20 kpc}$  encountering  different  spiral arms each one within a characteristic and opposite magnetic field whose coherence length  $ l_c \simeq{2 kpc}$ for He nuclei lead to:
 $$\delta_{rm-He} \simeq{16^\circ}\cdot \frac{Z}{Z_{He^2}} \cdot (\frac{6\cdot10^{19}eV}{E_{CR}})(\frac{B}{3\cdot \mu G})\sqrt{\frac{L}{20 kpc}} \sqrt{\frac{l_c}{2 kpc}}$$  This angle do fit the observed angles along UHECR clustering along Cen A for He and Be. Moreover the multiplet twenty EeV bending , if just He-like nuclei will be bent by a factor three times larger ($48^{o}$). If the multiplet at twenty EeV are made by fragments as Deuterium (or proton) ones, their bending will be only $\frac{3}{2}$ as large:
   $\delta_{rm-D}= \delta_{rm-p} \simeq{24^\circ}\cdot \frac{Z}{Z_{D^1}} \cdot (\frac{2\cdot10^{19}eV}{E_{CR}})(\frac{B}{3\cdot \mu G})\sqrt{\frac{L}{20 kpc}} \sqrt{\frac{l_c}{2 kpc}}$.  Such  bending angles are well comparable with the observed twenty EeV multiplet angle spread in narrow area as in figure see Fig.\ref{1r} and as it has foreseen since two and half years \cite{Fargion09a} up to recently \cite{Fargion2011}. Let us remind that our He-like UHECR do fit the AUGER and the HIRES composition traces. The He secondaries are splitting in two (or a fourth) energy fragments along Cen A tail (see Fig.\ref{1r}) within the  dotted circle around Cen A containing the two (of three) multiplet cores (see Fig.\ref{1r}) has a radius as small as $7.5^{o}$,  it extend in an area that is as smaller as  $180$ square degrees, well below  $1\% $ of the observation AUGER sky (see Fig.\ref{1r}). The probability that two among three multiplet sources fall inside this small area is offered by the binomial distribution: $ P (3,2) = \frac{3!}{2!} \cdot (10^{-2})^{2} \cdot \frac{99}{100}\simeq 3 \cdot 10^{-4}$. Moreover the same twin tail of the multiplet events are aligned almost exactly $\pm 0.1 $ rad along UHECR train. Therefore the UHECR  multiplet and alignment  has an a priori  probability to occur  as low as $ P (3,2) \simeq 3 \cdot 10^{-5}$ following a foreseen signature \cite{Fargion09a} ,\cite{Fargion2011}.   If contrary to present paper UHECR are mostly extragalactic nucleons they must respectively be leading to EeVs cosmogenic neutrino signals (if nucleon) at the edge of detection  by GZK cut off \cite{Greisen:1966jv}. Among neutrinos $\nu$,  muons ones $\nu_{\mu}$, are  deeply polluted by a rich atmospheric  component. At tens TeV-PeV up to EeV  $\nu_{\tau}$ neutrino might be nevertheless a clean signal of  UHECR-neutrino associated astronomy\cite{FarTau}. Their tau birth in ice  may shine as a double bangs (disentangled above PeV)  \cite{Learned}. UHE tau, born inside the  Earth or mountain, while escaping  in air  may lead, by decay in flight, to  loud, amplified  tau-airshowers \cite{Fargion1999},\cite{FarTau} to be observed in AUGER or TA in a few years\cite{FarTau},\cite{Auger-01},\cite{Feng02},\cite{Auger07} if UHECR are mostly nucleons.
\section{Conclusions: UHECR by He and $Ni^{56}$ , $Co^{60}$ radioactive nuclei? }
  The correlation of UHECR with Cen A, the absence of Virgo, the hint of correlation with  Vela and  with galactic TeV anisotropy, might be in part solved by an extragalactic lightest nuclei, mainly He, for Cen A , (see Fig.\ref{1r}) and mostly by heavy radioactive nuclei for other sources mostly in our own galaxy (see Fig.\ref{fig4-5}).  A partial confirm is the predicted \cite{Fargion09b} ,\cite{Fargion2011} and observed \cite{Auger11} multiplet clustering (as Deuterium or proton fragments) at half UHECR edge energy aligned  (see Fig.\ref{1r}): He like UHECR  maybe bent by a characteristic angle as large as  $\delta_{rm-He}  \simeq 16^\circ$; expected  proton or D fragments multiplet along tails spread  at $\delta_{rm-p}  \simeq 32^\circ$ \cite{Fargion2011},\cite{Auger11}. Moreover as shown  here, UHECR Ni,Co  and heavier nuclei too may be deflected by $\delta_{rm-Ni}  \simeq 18,7^{o}$ for nearby Vela,  $\delta_{rm-Ni}  \simeq  128^{o}$ (or less) for Crab, Gum Nebulae at few kpc away, forcing UHECR  toward nearby TeV inhomogeneities. Such TeV gamma anisotropy is made by boosted UHECR decay at hundred keV gamma and beta positrons shining to us at tens TeV.  In conclusion therefore we foresee  analogous UHECR traces around Cygnus and Cas A (as in TeV map) in growing future TA map. We foresee also crowding and clustering around ($\mp 18^{o}$) the galactic plane for similar arguments, but not much in galactic center where the stronger magnetic field may spray away the UHECR signals around an external corona ($\mp 20^{o}$). The UHECR radioactive beta decay in flight  may spread as  gamma TeV traces as well as $\nu$  TeVs-PeVs anisotropy whose muon signal is  sink into dominant atmospheric muon $\nu_{\mu}$ noise. A new spread $\nu_{\tau}$ neutrino astronomy, noise free,  related to astronomical (parasite oscillated) tau neutrino and its boosted tau (\emph{mini-double bang}  within a 5-50 meter size) in Deep Core  or Antares may reveal hundred TeV tau decay (seeing  similar PeVs ones in ICECUBE \cite{Learned}). Also Tau airshowers may rise in Cherenkov beamed air-showers. \cite{Fargion1999}, \cite{FarTau} or fluorescence telescopes at higher energies \cite{FarTau},\cite{Feng02},\cite{Auger08}.  The discover of such expected  Neutrino astronomy  may shed additional light on the UHECR nature, origination  and mass composition, while  opening our eyes to mysterious  UHECR sources. Also future gamma TeV maps and UHECR multiplet correlation as foreseen may  lead to a more conclusive fit of this unsolved, century old, cosmic ray puzzle. \\

\end{document}